\documentclass[aps,twocolumn,superscriptaddress,prl]{revtex4}
\usepackage{amssymb}
\usepackage{amsmath}
\usepackage{graphicx}

\begin{document}
\title{Effect of Exchange Interaction on Spin Dephasing in a Double Quantum Dot}
\author{E.\ A.\ Laird}
\affiliation{Department of Physics, Harvard University, Cambridge, Massachusetts
02138, USA}
\author{J.\ R.\ Petta}
\affiliation{Department of Physics, Harvard University, Cambridge, Massachusetts
02138, USA}
\author{A.\ C.\ Johnson}
\affiliation{Department of Physics, Harvard University, Cambridge, Massachusetts
02138, USA}
\author{C.\ M.\ Marcus}
\affiliation{Department of Physics, Harvard University, Cambridge, Massachusetts
02138, USA}
\author{A.\ Yacoby}
\affiliation{Department of Condensed Matter Physics, Weizmann Institute of
Science, Rehovot 76100, Israel}
\author{M.\ P.\ Hanson}
\affiliation{Materials Department, University of California at Santa Barbara,
Santa Barbara, California 93106, USA}
\author{A.\ C.\ Gossard}
\affiliation{Materials Department, University of California at Santa Barbara,
Santa Barbara, California 93106, USA}
\date{16 June 2006}

\begin{abstract} We measure singlet-triplet dephasing in a two-electron double
quantum dot in the presence of an exchange interaction which can be electrically
tuned from much smaller to much larger than the hyperfine energy. Saturation of
dephasing and damped oscillations of the spin correlator as a function of time
are observed when the two interaction strengths are comparable.  Both features
of the data are compared with predictions from a quasistatic model of the
hyperfine field.
\end{abstract}
\maketitle Implementing quantum information processing in solid-state
circuitry is an enticing experimental goal, offering the possibility of tunable
device parameters and
straightforward scaling. However, realization will require
control over the strong environmental decoherence typical of solid-state
systems. An attractive candidate system uses electron spin as the holder of
quantum information \cite{LossDiVincenzo, Taylor2005}.  In III-V semiconductor quantum
dots, where the highest degree of
spin control has been achieved~\cite{Fujisawa2002, ElzermanSingleShot,
BrackerOpticalSpin,
BraunOpticalSpin, HansonSingleShot, JohnsonT1, PettaT2}, the dominant decoherence mechanism is hyperfine interaction with the lattice
nuclei~\cite{hyperfinetheory}. A recent experiment
\cite{PettaT2} studied this decoherence in a qubit encoded in a pair of spins
\cite{LevySingletTriplet}. In this situation, the dynamics are governed by two
competing effects: the hyperfine interaction, which tends to mix the singlet and
triplet basis states, and exchange, which tends to preserve them.

The interplay of hyperfine and exchange effects has been studied recently via
spin-blockaded transport in two double-dot systems
\cite{onooscillations,koppensnuclei}. Oscillations and bistability \cite{onooscillations}
of the leakage current, as well as suppression of mixing with stronger exchange \cite{koppensnuclei}
were observed.  The topic also has a
long history in physical chemistry: recombination of a radical pair created in a
triplet state proceeds significantly faster for radicals containing isotopes
whose nuclei carry spin \cite{nucchemreviews}. By lifting the singlet-triplet
degeneracy, the exchange interaction suppresses spin transitions; its strength
can be deduced from the magnetic field dependence of the recombination rate
\cite{StaerkTarasov}. However, exchange is difficult to
tune \emph{in situ} in chemical systems.

\begin{figure}[b!]
\vspace{-0.3 cm} \centering \label{fig:fig1}
\includegraphics[width=3in]{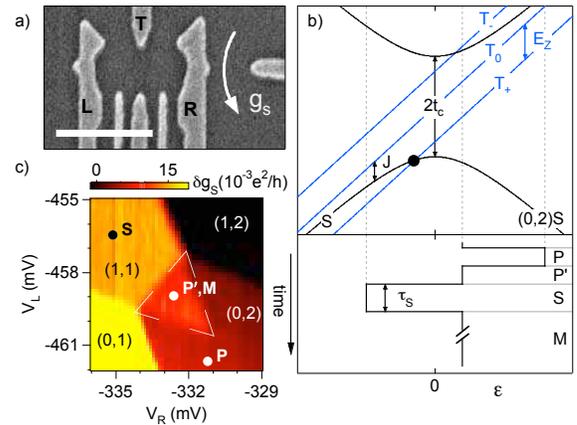}
\vspace{-0.3 cm} \caption{\footnotesize{(Color online) (a)
Micrograph of a device with the same gate design as the one
measured (Scale bar = 500 nm.) Voltages applied to gates L and R
are adjust the double dot detuning, $\epsilon$. Gate T sets the
inter-dot tunnel coupling. The conductance $g_\mathrm{s}$ of a
nearby sensor quantum point contact monitors the average
occupation of each dot. (b) Upper panel: Level diagram for the
double dot near the (1,1)-(0,2) transition ($\epsilon=0$) plotted
versus $\epsilon$.  Exchange ($J$) and Zeeman ($E_\mathrm{Z}$)
energies are indicated. $\bullet$ denotes the S-T$_+$ degeneracy.
Labels ($m,n$) denote the occupancies of the left and right dot
respectively. Lower panel: The prepare~(P, P$^{\prime}$) -
separate~(S) - measure~(M) pulse scheme. $\sim$90\% of the cycle
is spent in M.  (c)~$g_\mathrm{s}$ close to the (1,1)-(0,2)
transition during application of pulses, showing the pulse
triangle (marked) and the positions of points P, P$^{\prime}$, S
and M.  A background plane has been subtracted.}}
\end{figure}

In this Letter, singlet correlations between two separated
electrons in a GaAs double dot system are measured as a function
of a gate-voltage tunable exchange $J$ and as a function of time
$\tau_\mathrm{S}$ following the preparation of an initial singlet.
This study gives insight into the interplay of local hyperfine
interactions and exchange in a highly controllable quantum system.
We measure the probability $P_\mathrm{S}(\tau_\mathrm{S})$ that an
initial singlet will be detected as a singlet after time
$\tau_\mathrm{S}$ for $J$ ranging from much smaller than to much
greater than the rms hyperfine interaction strength in each dot,
$E_{\mathrm{nuc}}$.  When $J\ll E_\mathrm{nuc}$, we find that
$P_\mathrm{S}$ decays on a timescale $T_\mathrm{2}^{*}\equiv
\hbar/E_{\mathrm{nuc}}= 14$ ns.  In the opposite limit where
exchange dominates, $J\gg E_\mathrm{nuc}$, we find that singlet
correlations are substantially preserved over hundreds of ns. In
the intermediate regime, where $J\sim E_\mathrm{nuc}$, we observe
oscillations in $P_\mathrm{S}$ that depend on the ratio
$E_\mathrm{nuc}/J$.  Our results show that a finite exchange
energy can be used to extend spin correlations for times well
beyond $T_\mathrm{2}^{*}$.

 These observations are in reasonable agreement with recent theory,
which predicts a
singlet probability (assuming perfect readout)
$P^{0}_\mathrm{S}(\tau_\mathrm{S})$ that exhibits damped oscillations as a
function of time and a long-time saturation that depends solely on the
ratio $E_\mathrm{nuc}/J$ \cite{CoishLoss}.  To compare experiment and theory quantitatively we
introduce an empirical visibility, $V$, to account for readout inefficiency,
$P_\mathrm{S}(\tau_\mathrm{S}) =
1-V(1-P^{0}_\mathrm{S}(\tau_\mathrm{S}))$.

The device used in the experiment, shown in Fig.\ 1(a), is
fabricated on a GaAs/Al$_\mathrm{0.3}$Ga$_\mathrm{0.7}$As
heterostructure with a two-dimensional electron gas (density
$2\times10^{15}$~m$^{-2}$, mobility $20$ m$^2$/Vs)  100 nm below
the surface. Ti/Au top gates define a few-electron double quantum
dot. The inter-dot tunnel coupling $t_c$ and (0,2)-(1,1) detuning
$\epsilon$ are also separately tunable. A charge-sensing quantum
point contact with conductance $g_\mathrm{s} \sim 0.2e^2/h$ allows
the occupancy of each dot to be separately measured
\cite{FieldSensing, ElzermanSensing}. We monitor $g_\mathrm{s}$
using a lock-in amplifier with a 1~nA current bias at 335~Hz, with
a 30~ms time constant.

Measurements were made in a dilution refrigerator at electron temperature
$T_\mathrm{e}\approx100$ mK measured from the width of the (1,1)-(0,2)
transition \cite{DiCarlo}.  Gates L and R (see Fig.\ 1) were connected via
filtered coaxial lines to the outputs of a Tektronix AWG520.  We report
measurements for two settings of tunneling strength, controlled using voltages
on gate T and measured from the width of the (1,1)-(0,2) transition:
$t_c\approx23$ $\mu$eV (``large $t_c$") and $t_c<9$~$\mu$eV (``small $t_c$")
\cite{DiCarlo}. Except where stated,  measurements were made in a
perpendicular magnetic field of 200 mT, corresponding to a Zeeman energy
$E_\mathrm{Z} = 5$~$\mu$eV $\gg E_\mathrm{nuc}$.

Figure 1(b) shows the relevant energy levels near the (1,1)-(0,2)
charge transition as a
function of energy detuning $\epsilon$ between these charge
states. With $t_c$=0, the (1,1) singlet S and $m_s=0$ triplet
T$_0$ are degenerate; the $m_s=\pm1$ triplets T$_\pm$ are split
off in energy from T$_0$ by $\mp E_Z$. Finite $t_c$ leads to
hybridization of the (0,2) and (1,1) singlets, inducing an
exchange splitting $J$ between S and T$_0$. The (0,2) triplet (not
shown) is split off by the much larger intra-dot exchange energy
$J_{(0,2)} \sim 600 $ $\mu$eV \cite{JohnsonSpinBlockade} and is
inaccessible. Rapid mixing due to hyperfine interaction occurs
between states whose energies differ by less than $
E_\mathrm{nuc}$. This occurs at large negative $\epsilon$ (lower
left of Fig.\ 1(b)), where S and T$_0$ mix, and at
$J(\epsilon)=E_\mathrm{Z}$ (black dot in Fig.\ 1(b)), where S and
T$_+$ mix.

\begin{figure}[t!]
\center \label{fig:fig2}
\includegraphics[width=2.99in]{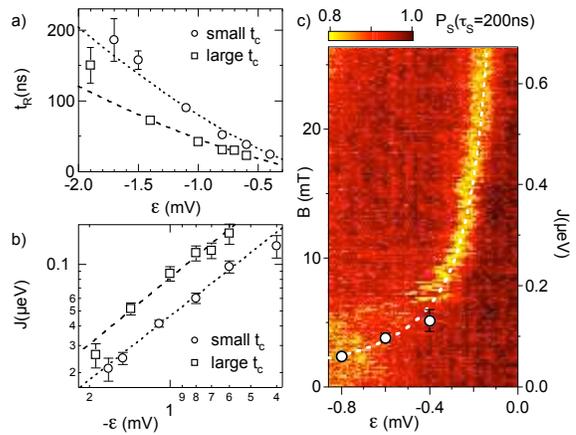}
\vspace{-0.3 cm} \caption{\footnotesize{(Color online) (a) Period
$t_R$ of first Rabi oscillation versus exchange point detuning for
small and large tunnel coupling. (b) Exchange energy as a function
of detuning, deduced from the data in (a), together with empirical
power-law fits $J \propto |\epsilon|^{-1.4 \pm 0.1}$.  $t_R$
corresponding to the fits is shown as curves in (a). (c) Color
scale plot of $P_\mathrm{S}$ as a function of S-point detuning and
magnetic field $B$ obtained using the pulse sequence in Fig. 1(b).
The bright band indicates rapid decoherence where $J=g
\mu_\mathrm{B}B$.  The white points and the dashed line are the
same data and fits plotted in (b).}}
\end{figure}

A cycle of gate configurations is used to prepare and measure
two-electron spin states \cite{PettaT2}, as illustrated in Fig.\
1(b). A 200~ns preparation step (denoted P in Fig.\ 1) configures
the dot in (0,2) at a position where the series (0,2)T$\rightarrow
$(0,1)$\rightarrow $(0,2)S is energetically allowed and occurs
rapidly, giving efficient initialization to a singlet. The gates
then shift (waiting 200~ns at P$^\prime$ to reduce pulse
overshoot) to a separation point (S) in (1,1) for a time
$\tau_\mathrm{S}$ during which singlet-triplet evolution occurs.
Finally, the gates are set to the measurement point~(M) for
$\tau_\mathrm{M}=5$~$\mu$s, for spin-to-charge conversion. Inside
the pulse triangle marked in Fig.\ 1(c), the triplet states will
remain in (1,1) over the measurement time $\tau_\mathrm{M}$
\cite{JohnsonT1,pettaT1}. Since $\sim$90\% of the pulse cycle is
spent at M, the relatively slow measurement of the sensor
$g_\mathrm{s}$ gives a time-averaged charge configuration at the M
point. This signal is calibrated to
give a
 singlet state probability $P_\mathrm{S}(\tau_\mathrm{S})$ by
comparing values within the pulse triangle with values within
(1,1) (which defines  $P_\mathrm{S}=0$)
and within (0,2) outside the pulse triangle (which defines $P_\mathrm{S}=1$). 

We first measure $J(\epsilon)$, $E_\mathrm{nuc}$, and $V$ at two
values of $t_\mathrm{c}$, allowing the saturation probability
$P_\mathrm{S}(\infty )$ to be measured as a function of $J$.  This
saturation probability is found to depend on the ratio
$E_\mathrm{nuc}/J$ approximately as predicted by theory
\cite{CoishLoss}.  We then measure the time evolution
$P_\mathrm{S}(\tau_\mathrm{S})$, which shows damped oscillations,
also in reasonable agreement with theory \cite{CoishLoss}.
$J(\epsilon)$ is measured using the Rabi (or Larmor) sequence
described in Ref.~\cite{PettaT2}, in which an adiabatic (compared
with $E_\mathrm{nuc}$) ramp over 1~$\mu$s to (1,1) is used to
prepare and measure the electron spin state in the
$\{\left|\uparrow\downarrow\right\rangle,
\left|\downarrow\uparrow\right\rangle\}$ basis. An exchange pulse
produces coherent rotations with a period $t_\mathrm{R}$ (shown in
Fig.~2(a)) from which we deduce the exchange coupling
$J(\epsilon)= h/t_R$ \cite{footnote2}.  Values of $J(\epsilon)$
for small and large $t_c$ are shown in Fig.~2(b), along with a fit
to an empirical power-law form $J \propto \epsilon^{-\alpha}$,
giving $\alpha \sim 1.4$ \cite{footnote3}. In Fig.~2(c), these
values of $J(\epsilon)$ are compared with the results of an
alternative method in which rapid dephasing at the S-T$_+$
degeneracy produces a dip in $P_\mathrm{S}$ when the value of
$\epsilon$ at the S point satisfies $J(\epsilon)=E_\mathrm{Z}$.
$J(\epsilon)$ can then be measured from a knowledge of the field,
using $E_\mathrm{Z} = g\mu_\mathrm{B}B$ where $\mu_\mathrm{B}$ is
the Bohr magneton, and taking the value $g = -0.44$, measured
(using an in-plane field) in a different quantum dot device on
made from the same wafer \cite{ZumbuhlFED}.  $J(\epsilon)$
measured by this technique is in qualitative agreement with the
power-law derived from Fig.~2(b); discrepancies may be due to an
anisotropic $g$-factor,  nuclear polarization effects, or may
indicate a dependence of $J(\epsilon)$ on field. Since the first
method more closely matches the conditions under which data in the
rest of the paper was taken and is more precise in the range of
$J$ of interest, we henceforth take $J(\epsilon)$
from Fig.~2(b).

Parameters $E_\mathrm{nuc}$ and $V$ are extracted from $P_\mathrm{S}(\tau_\mathrm{S})$
measured for the S-point at large negative $\epsilon$, where $J\ll
E_\mathrm{nuc}$. In this regime the initial singlet evolves into an equal
mixture of singlet and triplet with
characteristic time $h/E_\mathrm{nuc}$. $P_\mathrm{S}(\tau_{S})$ for small and
large $t_c$ (shown in the insets of Fig.~3) are fit to the form
for $P^{0}_\mathrm{S}(\tau_\mathrm{S})$ given in \cite{CoishLoss},
with fit parameters $E_\mathrm{nuc}=45\pm3$ neV
($47\pm 4$~neV), corresponding to hyperfine fields of $\sim$ 1.8 mT, and $V=0.53\pm0.06\
(0.46\pm0.06)$ for small (large) $t_c$ \cite{footnote4}.
The fit function $P^{0}_\mathrm{S}(\tau_\mathrm{S})$ depends on $J$ at this detuning, which is too small to measure
directly. Instead, $J(\epsilon)$ is extrapolated using the power-law from Fig.\
2b; however, the fit parameters are essentially independent of details of the
extrapolation, and, for example, are within the error bars for the extrapolation
$J \propto |\epsilon|^{-1}$ as well as $J=0$.

$P^{0}_\mathrm{S}
(\tau_\mathrm{S})$ is expected to show a range of interesting
behavior
depending on the relative magnitudes of $J$ and $E_\mathrm{nuc}$ \cite{CoishLoss}:  In the
limit $J=0$, $P^{0}_\mathrm{S}(\tau_{S}\rightarrow
\infty)$ rapidly saturates to 1/2. As $J$ is increased, hyperfine dephasing
becomes less effective, with $P^{0}_\mathrm{S}(\infty)$
saturating at progressively higher values, approaching unity when $J \gg
E_\mathrm{nuc}$, and following a universal function of $E_\mathrm{nuc}/J.$  As a
function of
$\tau_\mathrm{S}$, $P^{0}_\mathrm{S}(\tau_{S})$ is predicted to undergo damped
oscillations, which when plotted versus $\tau_\mathrm{S}J$ follow another
universal function of $E_\mathrm{nuc}/J$ and exhibit a universal phase shift of
$3\pi/4$ at large $\tau_\mathrm{S}J$.

Knowing $J(\epsilon)$ and $E_\mathrm{nuc}$ allows the
long-time ($\tau_\mathrm{S}\gg h/J$) saturation of the measured
$P_\mathrm{S}$ to be compared with theory \cite{CoishLoss}.
We set $\tau_\mathrm{S}=400$ ns and sweep the position of the S-point. For small
and large $t_\mathrm{c}$, $P_\mathrm{S}(400$ ns) is plotted in Fig.\ 3 as a
function of $E_\mathrm{nuc}/J$, where $E_\mathrm{nuc}$ is obtained from the fits
described above and $J(\epsilon)$ are taken from Fig.~2. At the
most negative detunings (in the regions marked by gray bars in Fig.~3) $J$ is
too small to be measured by either Rabi period or S-T$_{+}$ degeneracy methods;
instead, $J(\epsilon)$ is found by extrapolating the power-law fits. As above,
agreement
with theory (discussed below) is insensitive to the details of the extrapolation, as shown by the dashed
lines in Fig. 2.

\begin{figure}[t!]
\center \label{fig:fig3}
\includegraphics[width=2.4in]{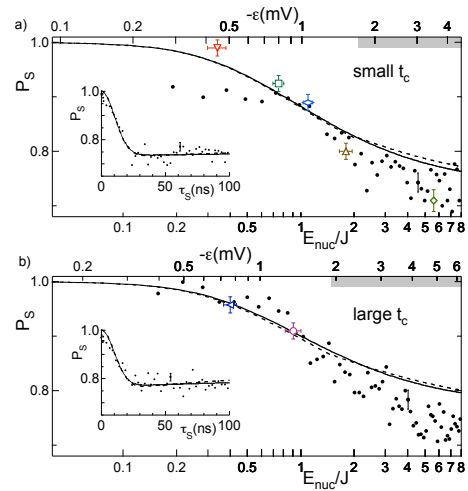}
\vspace{-0.3 cm} \caption{\footnotesize{(Color online) (a) Inset:
$P_\mathrm{S}(\tau_\mathrm{S})$ for small $t_c$ and $\epsilon$=
-5.5 mV, with fit (see text) giving $E_\mathrm{nuc}$=45$\pm$~3~neV
and $V$=0.53$\pm$0.06. Main panel: Measured
$P_\mathrm{S}$$(\tau_\mathrm{S}$=400 ns) (points) plotted against
$E_\mathrm{nuc}/J$.  Open symbols correspond to $P_\mathrm{S}$ in
the traces of Fig.\ 4(a) at the largest $\tau_\mathrm{S}$ measured
for each $\epsilon.$ Curve shows theoretical dependence (from
\cite{CoishLoss}) of
$P_\mathrm{S}$$(\tau_\mathrm{S}\rightarrow\infty)$ on
$E_\mathrm{nuc}/J$, taking into account the measurement fidelity
deduced from the inset.  The gray bar along the top axis indicates
the region where $J(\epsilon)$ is extrapolated (see text).  Dashed
lines indicate the theoretical predictions (plotted as functions
of $\epsilon$) if an alternative extrapolation $J\propto
|\epsilon|^{-1}$ is chosen in this region.  (b)Large $t_c$ data.
The fit to the inset gives $E_\mathrm{nuc}=47\pm 4$ neV and
$V=0.46\pm0.06$, from which the theoretical saturation
$P_\mathrm{S}$ (curve in main panel) is calculated. Open symbols
correspond to the large-$\tau_\mathrm{S}$ values in Fig.~4(b).
Error bars on the filled symbols shows the uncertainty in
$P_\mathrm{S}$ arising from charge noise in the sensing point
contact.}}
\end{figure}

The long-$\tau_{S}$ $P_\mathrm{S}$ data shown in Fig.\ 3 agrees fairly well with the saturation values predicted from
\cite{CoishLoss}, taking into account the visibility (assumed independent of $\epsilon$) obtained from the insets. In
particular, $P_\mathrm{S}$ has the same dependence on $E_\mathrm{nuc}/J$ at both values of $t_\mathrm{c}$ measured, even
though the function $J(\epsilon)$ depends on $t_\mathrm{c}$. $P_\mathrm{S}$ is up to $\sim0.06$ smaller than predicted
at the largest detunings; both cotunneling and nuclear decorrelation over the duration of the
separation pulse tend to equalize singlet and triplet occupations, although it is unclear whether they are the cause
of this effect.

We next investigate the time dependence of
$P_\mathrm{S}(\tau_\mathrm{S})$ at finite $J$.  For five (two)
S-point detunings at small (large) $t_c$,
$P_\mathrm{S}(\tau_\mathrm{S})$ was measured out to
$\tau_\mathrm{S}J/\hbar\approx15$. The results are shown in Fig.\
4, together with the predicted time evolution
from~\cite{CoishLoss} with values for $V$ and $E_\mathrm{nuc}$
taken from fits shown in the insets of Fig.~3.  Because
$P_\mathrm{S}$ remains close to unity, these data are particularly
sensitive to calibration imperfections caused by quantum point
contact nonlinearities and noise in the calibration data, whose
effect to lowest order is to shift the data vertically. Traces in
Fig.\ 4 are therefore shifted vertically to satisfy the constraint
$P_\mathrm{S}(\tau_\mathrm{S}=0)=1$.  In no case was this greater
than $\pm$0.05. Here and in Fig.~3, the error bars reflect
uncertainty in $P_\mathrm{S}$ from charge noise in the sensing
point contact; additional scatter in the data may be due to long
nuclear correlation times \cite{PettaT2, koppensnuclei}.

\begin{figure}[!t]
\center \label{fig:fig4}
\includegraphics[width=2.4in]{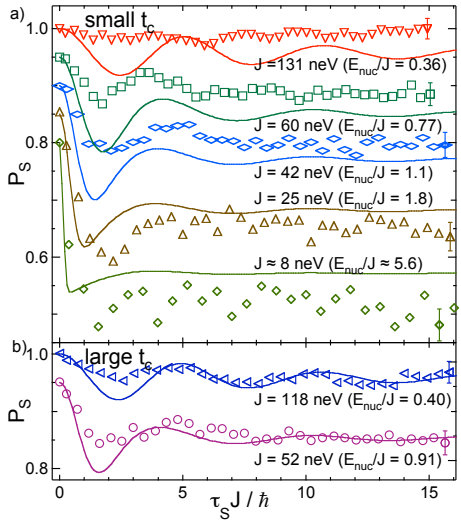}
\vspace{-0.3 cm} \caption{\footnotesize{(Color online) (a)
Symbols: Experimental $P_\mathrm{S}(\tau_\mathrm{S})$ at small
$t_c$ for various J, plotted as a function of
$\tau_\mathrm{S}J/\hbar$. Curves: Predictions
from~\cite{CoishLoss} using $E_\mathrm{nuc}$ and $V$ fit from
Fig.\ 3(a). Adjacent traces after the first are offset by 0.05 for
clarity. (b) Corresponding data and theory for large $t_c$.  Lower
trace is offset by 0.05 for clarity.  Error bars reflect the
contribution of sensor charge noise.}}
\end{figure}

Damped oscillations are observed as predicted in \cite{CoishLoss}; however, even
after taking account the empirical visibility factor, the amplitude of the
oscillations is less than expected. This is likely due to the
finite rise time of the separation pulse and to switching noise, which make
each trace effectively an average over a range of $J$ values.  Where the amplitude is large enough for the period and
phase of the oscillations to be made out, these approximately match the predictions of \cite{CoishLoss}, although with
two significant departures: The
topmost trace, with smallest $E_\mathrm{nuc}/J$, does not show clear oscillations, and the expected shift of the first
minimum to smaller $\tau_\mathrm{s}J$ at intermediate $J$ is not observed.  We do not understand the origin of these
effects.  The amplitude of the oscillations falls off too rapidly for the expected $3\pi/4$ phase shift at large
$\tau_\mathrm{S} J$ to be visible.  Similar
oscillations of $P_\mathrm{S}$
are predicted close to the S-T$_+$ degeneracy with a characteristic frequency $\sim\Delta=J-E_\mathrm{Z}$.  We have
searched for these oscillations but do not observe them.  We believe the reason
for this is that $\Delta$ varies much more rapidly with $\epsilon$ in this
region than $J$ does at the S-T$ _0$ near-degeneracy; the oscillations are
therefore washed out by switching noise and pulse overshoot.

In summary, after including the measured readout efficiency, we
find that the singlet correlator shows damped oscillations as a
function of time and saturates at a value that depends only on
$E_\mathrm{nuc}/J$.  Both these features are qualitatively as
expected from theory \cite{CoishLoss}; some of the departures from
expected behavior may be qualitatively accounted for by
cotunneling and nuclear decorrelation (which tend to equalize
singlet and triplet probabilities at long times), and charge noise
(which tends to smear out the oscillations seen in Fig.~4.)

We acknowledge useful discussions with W. Coish, H.~A. Engel, D. Loss, M. Lukin,
J. M. Taylor. This work was supported by DARPA-QuIST and the ARO/ARDA/DTO STIC
program.

\small
\bibliographystyle{prl}

\begin{thebibliography}{25}
\bibitem{footnote2} When $J\lesssim E_\mathrm{nuc}$, $J$ must be corrected
downwards slightly because precession in the nuclear field
enhances the average Rabi frequency.  The correction to $J$ never
exceeds~13\%.
\bibitem{footnote3} A simple level anticrossing with $t_{c}$ and $\epsilon$
independent would give $J \propto \epsilon^{-1}$.  The discrepancy may be due to
a detuning-dependent $t_{c}$.
\bibitem{footnote4} The dependence of $V$ on $t_{c}$
is not understood.  The effect of
alternative visibility parameters on the predictions of Figs. 3 and 4 is simply to scale them towards or away from $P_{S}=1$
\end{thebibliography}

\end{document}